\documentclass[reprint,amsmath,amssymb,aip,pof]{revtex4-1}
\usepackage{graphicx}% Include figure files
\usepackage{dcolumn}% Align table columns on decimal point
\usepackage{bm}% bold math
\usepackage{soul,color,amsmath}
\setstcolor{red}
\begin{document}

\preprint{Bhateja-Jain-2021}

\title{Self-similar velocity and solid fraction profiles in silos with eccentrically-located outlets}

\author{Ashish Bhateja}
\email{ashish@iitgoa.ac.in}
\affiliation{School of Mechanical Sciences, Indian Institute of Technology Goa, Ponda 403401, Goa, India}

\author{Sahaj Jain}
\altaffiliation[Now at ]{Department of Biomedical Engineering \& Mechanics, Virginia Tech, Blacksburg, USA}
\affiliation{School of Mechanical Sciences, Indian Institute of Technology Goa, Ponda 403401, Goa, India}

\date{\today}

%% Abstract
\begin{abstract}
We examine the gravity-induced flow of dry and cohesionless granular media through an outlet placed eccentrically in a planar silo, employing computations based on a soft-sphere discrete element method. The vertical velocity profiles, measured at the outlet, are self-similar when the eccentricity is given in terms of the gap ($s$) between the wall and the corner of the outlet nearest to the wall. On the other hand, the self-similarity of vertical velocity does not always hold for all eccentricities ($e$) given by the distance between the centers of an outlet and the silo base, which is a typical metric of eccentricity. Similar observations are noted for the profiles of solid fraction. For the former measure of eccentricity, the flow conditions are observed to be similar for different outlet sizes. In contrast, we observe, the latter leads to differing flow patterns for the highest eccentricity wherein the largest outlet touches the sidewall and the rest are located at a distance. This study establishes the importance of $s$ compared to $e$ from the viewpoint of the self-similarity of the vertical velocity and solid fraction profiles at the outlet, and generalizes the notion of the scaling of velocity and solid fraction reported by Janda \textit{et al.} [Phys. Rev. Lett. \textbf{108}, 248001 (2012)] in a silo with a centric exit to the one with eccentric granular discharge.
\end{abstract}
%%
%\pacs{81.05.Rm, 45.70. Mg, 45.70.-n}% PACS, the Physics and Astronomy
%\keywords{Discrete Element simulations}%Use showkeys class option if keyword display desired

\maketitle

%\tableofcontents

%\section{\label{sec:level1}First-level heading:\protect\\ The line
%break was forced \lowercase{via} \textbackslash\textbackslash}
% Introduction
\section{Introduction}
Silos are widely used equipment for handling granular materials in various industrial settings. Besides jamming, one of the prime concerns in such systems is obtaining a robust expression for estimating mass flow rate for various geometrical and material parameters\cite{beverloo1961,mankoc2007,janda2012,rubio2015}, requiring a thorough understanding of flow near the outlet. Such an insight is not only limited to the flow of discrete media in silos and hoppers, but it also enables us to understand other strikingly similar systems better, such as pedestrian dynamics\cite{helbing2000,pastor2015}. 

Most research investigations examining granular flow in silos focus on the arrangement where an outlet is placed at the center of the base\cite{nedderman1982,saleh2018}. Studies exploring granular flow through eccentric exits are relatively less\cite{saleh2018}. Sielamowicz and co-workers presented an empirical description of velocity for eccentric granular discharge considering scenarios wherein the filling and discharge locations are on the same\cite{sielamowicz2011} and opposite\cite{sielamowicz2010} sides of the central axis of the silo, with the outlets stationed close to the walls. A recent study by Maiti \textit{et al.}\cite{maiti2016a} examined the discharge of granular media through eccentrically-located apertures, including the one placed at the extreme position adjoining the wall in a silo. They extended the kinematic model of Nedderman and T\"{u}z\"{u}n\cite{nedderman1979} and validated theoretical predictions with experiments. A later work of Maiti and colleagues\cite{maiti2016b} in a silo reports an extensive experimental investigation focussing on variation in the flow pattern when an outlet is moved from the centric position to an eccentric location. 
%Interestingly, the study also showed profiles for the dimensionless vertical velocity at a fixed location vertically above the outlet for different eccentricities\cite{maiti2016b}.

The previous decade witnessed the appearance of many studies\cite{janda2012,zhou2015,rubio2017,madrid2017,gella2017,bhateja2020} reporting the self-similarity of velocity profiles at the centric outlet of a discharging silo, providing new avenues for deepening insights into the mechanics of granular flow near the outlet. Experiments of Janda \textit{et al.}\cite{janda2012} obtained the scaling of vertical velocity at the exit in  a quasi-two-dimensional silo draining under gravity, with the outlet size $D$ being the length scale. Subsequently, such scaling is reported, experimentally\cite{madrid2017,gella2017} and computationally\cite{zhou2015,rubio2017,bhateja2020}, in different granular systems. For example, Zhou \textit{et al.}\cite{zhou2015} reproduced the scaling in two-dimensional simulations, based on the contact dynamics method\cite{radjai2009}, for the polydisperse and bidisperse granular media. The experiments of Madrid \textit{et al.}\cite{madrid2017} demonstrated the scaling to hold for a bidisperse mixture with large size ratios. Discrete element simulations of Rubio-Largo \textit{et al.}\cite{rubio2017} and Bhateja\cite{bhateja2020} in three and two dimensions, respectively, obtained the said velocity scaling for a polydisperse granular assembly. Gella \textit{et al.}\cite{gella2017} extended the work of Janda \textit{et al.}\cite{janda2012} and reproduced the velocity scaling for a monodisperse granular system with a larger particle size. The studies mentioned above utilize silos with a flat base. The self-similarity of velocity profiles is also found to hold in hoppers discharging under gravity in recent experiments\cite{darias2020,mendez2021} and simulations\cite{mendez2021} based on Computational Fluid Dynamics (CFD). In addition to velocity, the self-similarity of solid fraction profiles at the exit has also been reported in various granular systems described before\cite{janda2012,zhou2015,rubio2017,gella2017,darias2020}.

Note that the aforestated investigations on the scaling of velocity and solid fraction are carried out in silos having centric apertures. Therefore, naturally, one could think of the appearance of self-similar velocity and solid-fraction profiles in silos having eccentrically-located outlets, especially those placed in proximity to the walls. This work, thus, aims to examine the self-similarity of such kinematic quantities at the exit stationed eccentrically at the base of a silo discharging under gravity. To this end, we utilize computations based on the discrete element method\cite{cundall1979,shafer1996,bkm2003}. The paper is organized as follows. Section~\ref{sec:setup} presents the computational methodology along with features of the granular system being considered. Results are presented and discussed in Sec.~\ref{sec:results}, followed by conclusions in Sec.~\ref{sec:conclude}.

% Computational procedure
\section{Computational technique and set-up}
\label{sec:setup}
Computations presented in this paper are based on a soft-sphere discrete element method\cite{cundall1979,bkm2003}. Particles are assumed to be circular, with mean diameter $d$ and mass density $\rho$. A polydispersity of $\pm 10\%$ is considered in particle size. Particles are considered to be dry and cohesionless. 

The forces considered on each particle are due to gravity and its contact with other particles and boundaries. The contact force between particles is computed using the linear spring-dashpot model\cite{zhang1996}. The normal and tangential components of the contact  force acting on particle $i$ due to particle $j$ are given by
\begin{eqnarray}
\bm{F}_n &=& -k_n \, \delta \, \bm{\hat{n}} - \gamma_n \, \bm{u}_n, \\
\bm{F}_t &=& - \gamma_t \, \bm{u}_t,
\end{eqnarray}
where $k_n$ is the normal spring stiffness, and $\gamma_n$ and $\gamma_t$ are the normal and tangential damping coefficients, respectively. In this study, we do not consider spring in the tangential direction. Particles $i$ and $j$, having radii $R_i$ and $R_j$, respectively, are said to be in contact when the overlap $\delta = (R_i + R_j - |\bm{r}_{ij}|)$ is equal to or greater than zero, where  $|\bm{r}_{ij}|$ gives the distance between the centers of particles and $\bm{r}_{ij} = \bm{r}_j - \bm{r}_i$. The unit vector $\bm{\hat{n}}=\bm{r}_{ij}/|\bm{r}_{ij}|$ directs along the line joining the centers of particle $i$ to particle $j$. The relative velocity $\bm{u}_{ij}$ of particle $i$ with respect to particle $j$ at the point of contact is given as $\bm{u}_{ij} = \bm{u}_i - \bm{u}_j + (R_i\, \bm{\omega}_i + R_j\, \bm{\omega}_j) \, \times \bm{\hat{n}}$, with $\bm{\omega}_i$ and $\bm{\omega}_j$ being the angular velocity of particle $i$ and $j$, respectively. The normal and tangential components of the relative velocity are $\bm{u}_n= (\bm{u}_{ij} \cdot \bm{\hat{n}})\,\bm{\hat{n}}$ and $\bm{u}_t = \bm{u}_{ij} - \bm{u}_n$. Considering the Coulomb's criterion, the magnitude of the tangential force is restricted to $\mu_p \, |\bm{F}_n|$, thus
\begin{equation}
\bm{F}_t = - \text{min} (\mu_p \, |\bm{F}_n| \bm{\hat{t}}, \gamma_t \, \bm{u}_t),
\end{equation}
where $\mu_p$ is the friction coefficient and $\bm{\hat{t}}=\bm{u}_t/|\bm{u}_t|$ is the unit vector along tangential direction. The damping coefficient $\gamma_n$ is related to the coefficient of normal restitution $e_n$ in the following manner for the linear spring-dashpot model\cite{bkm2003}
\begin{equation}
\gamma_n = 2 \,  \left(\frac{k_n \, m_e \, (\text{ln}\, e_n)^2} {\pi^2 + (\text{ln}\,e_n)^2}\right)^{1/2},
\label{Eq:cn}
\end{equation}
where $m_e=m_i m_j/(m_i + m_j)$ gives the effective mass of colliding particles $i$ and $j$, with $m_i$ being the mass of particle $i$.

The interaction between a particle and a wall is modelled using the same force scheme as particle-particle interactions. The equations of motion are integrated using the velocity-Verlet algorithm\cite{kruggel2008b}. All quantities reported here are nondimensionalized using the mean particle diameter $d$, mass density $\rho$, and acceleration due to gravity $g$. The following parameters, which are same for particle-particle and particle-wall interactions, are used in our computations: the normal spring stiffness $k_n=10^6$, friction coefficient $\mu_p=0.4$, coefficient of restitution $e_n=0.9$, and the time step for integration $\Delta t = 10^{-4}$. The normal and tangential damping constants are same, i.e., $\gamma_n=\gamma_t$. 

% Schematic and simulation snapshot
\begin{figure}[ht!]
\includegraphics[scale=0.46]{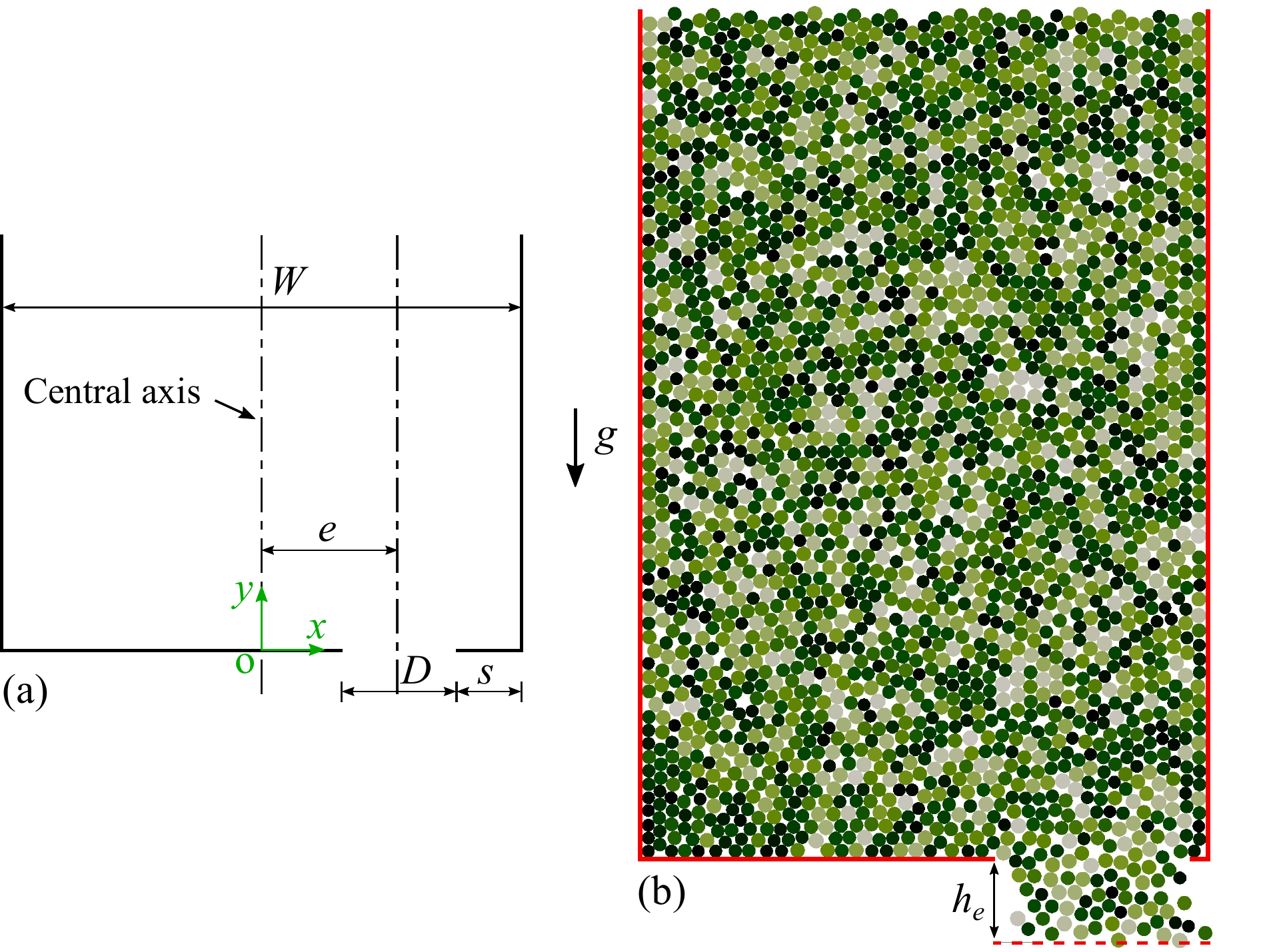}
\caption{(a) A schematic of the silo showing various dimensions. (b) A simulation snapshot depicting the flow of granular media through an eccentrically-located exit at the base of the silo. The direction of gravity $g$ and coordinate axes are also appropriately indicated.}
\label{fig:snapshot}
\end{figure}

We consider a two-dimensional silo. A schematic representation of the silo is shown in Fig.~\ref{fig:snapshot}, along with a snapshot illustrating the flow of grains through an outlet stationed eccentrically at the silo base. Grains crossing the vertical location $h_e$ below the outlet (see Fig.~\ref{fig:snapshot}(b)) are reinserted into the system at random horizontal positions on top with the same velocity (not shown for brevity). Doing so maintains the fill height of grains broadly to its initial value, i.e., $H \sim 305$. Number of grains corresponding to this fill height is $N=13200$. The silo width is $W=40$, and the outlet size $D$ is varied  between $9$ and $14$ in increments of one mean particle diameter. The outlet size is kept higher than six mean particle diameter in order to eliminate the possibility of jamming at the outlet\cite{janda2008,mankoc2007,kondic2014}. 

We place outlets right to the central axis as shown in Fig.~\ref{fig:snapshot}(a). There are two ways in which the eccentricity of an outlet is measured. In the first approach, it is given by the distance ($e$) between the centers of an outlet and the silo base. The second measure is the gap ($s$) between the right corner of an outlet and the right wall. It is clear from Fig.~\ref{fig:snapshot}(a) that $e$ and $s$ are related through $e = [(W-D)/2-s]$. In other words, $s$ changes when $e$ is same for all outlet sizes, and vice versa. In this work, $e$ is varied from 0 to 13 with unit increments, and $s$ ranges between 0 and 12 with an interval of one. Note that $s=0$ corresponds to the situation where an outlet touches the right wall.

All results reported here are obtained in the steady state. Averaging is done over 50 data sets. Each data set corresponds to a simulation beginning with a different configuration and averaged over 10000 snapshots recorded in an interval of 100 time steps. The flow fields are obtained by using the Heaviside function of coarse-graining\cite{goldhirsch2010} width $w=d$. For a bin centered at point $\bm{x}$, the mean velocity $\bm{v}(\bm{x})$ and solid fraction $\phi(\bm{x})$ are given by
\begin{eqnarray}
\bm{v} &=& \frac 1N_{b} \sum_{i=1}^{N_b} \bm{u}_{i}, \\ 
\phi &=& \sum_{i=1}^{N_b} A_i/A_b,
\end{eqnarray}
where $\bm{u}_i$ is the particle-wise instantaneous velocity, $A_i$ and $A_b$ are the particle and bin area, respectively, and $N_b$ is the number of particles located within $w/2$ distance from the bin center such that $|x_i-x_b|\leq w/2$ and $|y_i-y_b|\leq w/2$. Here, $\{x_i,y_i\}$ and $\{x_b,y_b\}$ are coordinates of the particle and bin centers, respectively. Keeping the area same for all bins, the aforesaid cut-off distances are appropriately considered for the bins centered on and within $w/2$ distance from the boundaries.

% Results and discussions
\section{Results and discussions}
\label{sec:results}
Let us now present results obtained from simulations. It has been verified that there is no effect of removing particles crossing the vertical location $h_e$ below the outlet for $h_e=6$ and above. Therefore, we consider $h_e=6$ for all outlet sizes and eccentricities. Note that the profiles of velocity and solid fraction reported here are measured at the outlet, i.e., at the vertical location $y=0$.

% Velocity
\subsection{Velocity}
We focus our attention on scaling of the vertical component of the mean velocity $v_y$, as what is done in earlier investigations\cite{janda2012,zhou2015,rubio2017,madrid2017,gella2017}. First, we examine the horizontal profiles of $v_y$ for all outlet sizes for the eccentricity ($e$) measured from the base center. Note that $e=0$ for the centric outlets. For brevity, the data for $e=0$ and $13$ are presented.
%%
% Figures for the velocity and its scaling for e=0
\begin{figure}[ht!]
\includegraphics[scale=0.5]{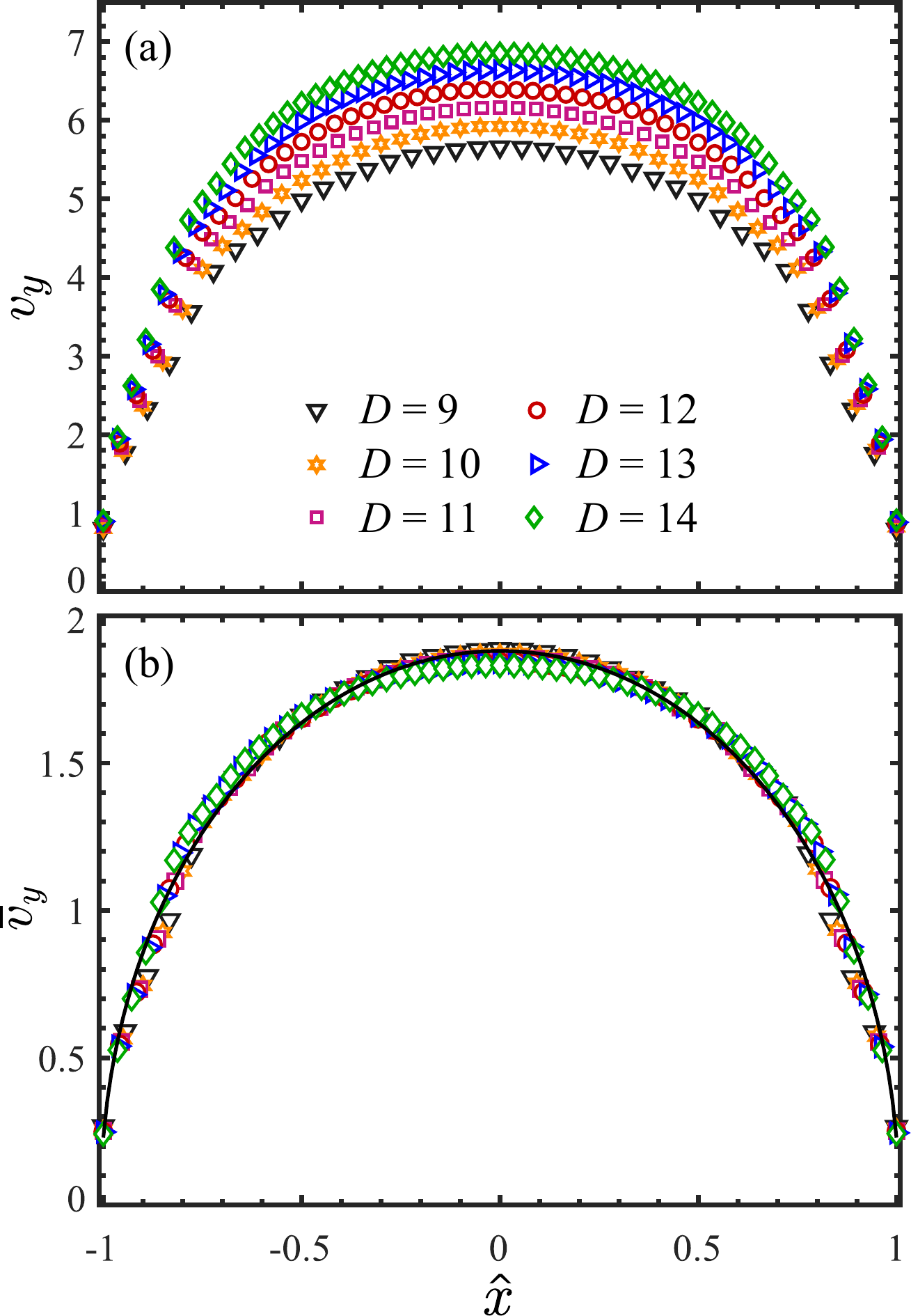}
\caption{Variation of (a) the vertical component of the mean velocity $v_y$ and (b) the scaled vertical velocity $\overline{v}_y$ with $\hat{x}=2(x-e)/D$ for the centric outlets ($e=0$). The black solid line in (b) represents Eq.~(\ref{eqn:scale1}) with fitting parameters mentioned in the text. Error bars, which are computed based on the standard error\cite{altman2005}, are not displayed as these are smaller than the markers. Legend for both plots is provided in (a).}
\label{fig:ve0}
\end{figure}
Figs.~\ref{fig:ve0}(a) and \ref{fig:ve0}(b) show the variation of $v_y$ and scaled velocity $\overline{v}_y=v_y/(gD)^{1/2}$ with $\hat{x}=2(x-e)/D$ for the centric apertures, respectively. Note that $\hat{x}=-1,1,$ and $0$ correspond to the left and right corners, and center of the outlet, respectively. As expected, the velocity profiles are symmetric about $\hat{x}=0$. For a given outlet size, velocity increases as one moves from the outlet corner to its center, where it achieves the maximum value, as shown in Fig.~\ref{fig:ve0}(a). The self-similarity of velocity profiles is demonstrated in Fig.~\ref{fig:ve0}(b) as the scaled velocity data for different $D$ superimpose, in striking agreement with previous computations\cite{zhou2015,rubio2017} and experiments\cite{janda2012,madrid2017,gella2017}. The following equation is fit to the scaled velocity data presented in Fig.~\ref{fig:ve0}(b) for $e=0$,
\begin{equation}
\overline{v}_y = \alpha_1 \, (1-\alpha_2 \, \hat{x}^2)^{\nu},
\label{eqn:scale1}
\end{equation}
with fitting parameters $\alpha_1 = 1.88, \alpha_2 = 0.98$ and $\nu=0.49$. The exponent $\nu$ matches very well with what is reported ($\nu = 0.5$) in the experimental investigation of Janda \textit{et al.}\cite{janda2012}, notwithstanding that they used spheres whereas we employ discs. The expression given by Eq.~(\ref{eqn:scale1}) is similar to what is introduced by Darias \textit{et al.}\cite{darias2020} for taking care of non-zero velocity at the outlet border in granular flow through a hopper. 

% Figures for the velocity and its scaling for e=13
\begin{figure}[ht!]
\includegraphics[scale=0.5]{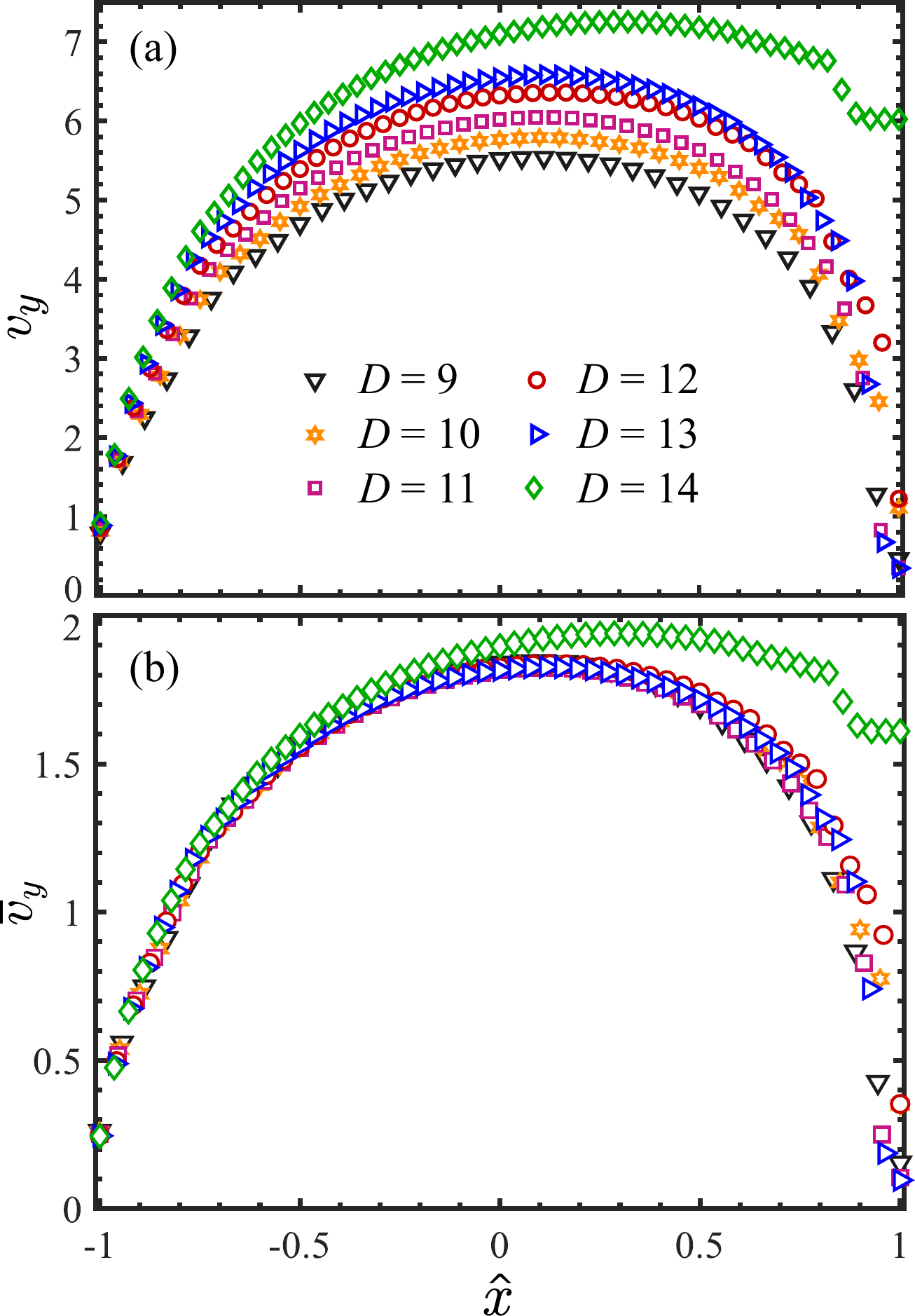}
\caption{Horizontal profiles of (a) the vertical velocity $v_y$ and (b) the scaled vertical velocity $\overline{v}_y$ for the outlets having eccentricity $e=13$. Error bars are not shown as these are smaller than the markers. Legend for both plots is given in (a).}
\label{fig:ve13}
\end{figure}

% Spatial distribution of the vertical velocity
\begin{figure}[ht!]
\includegraphics[scale=0.45]{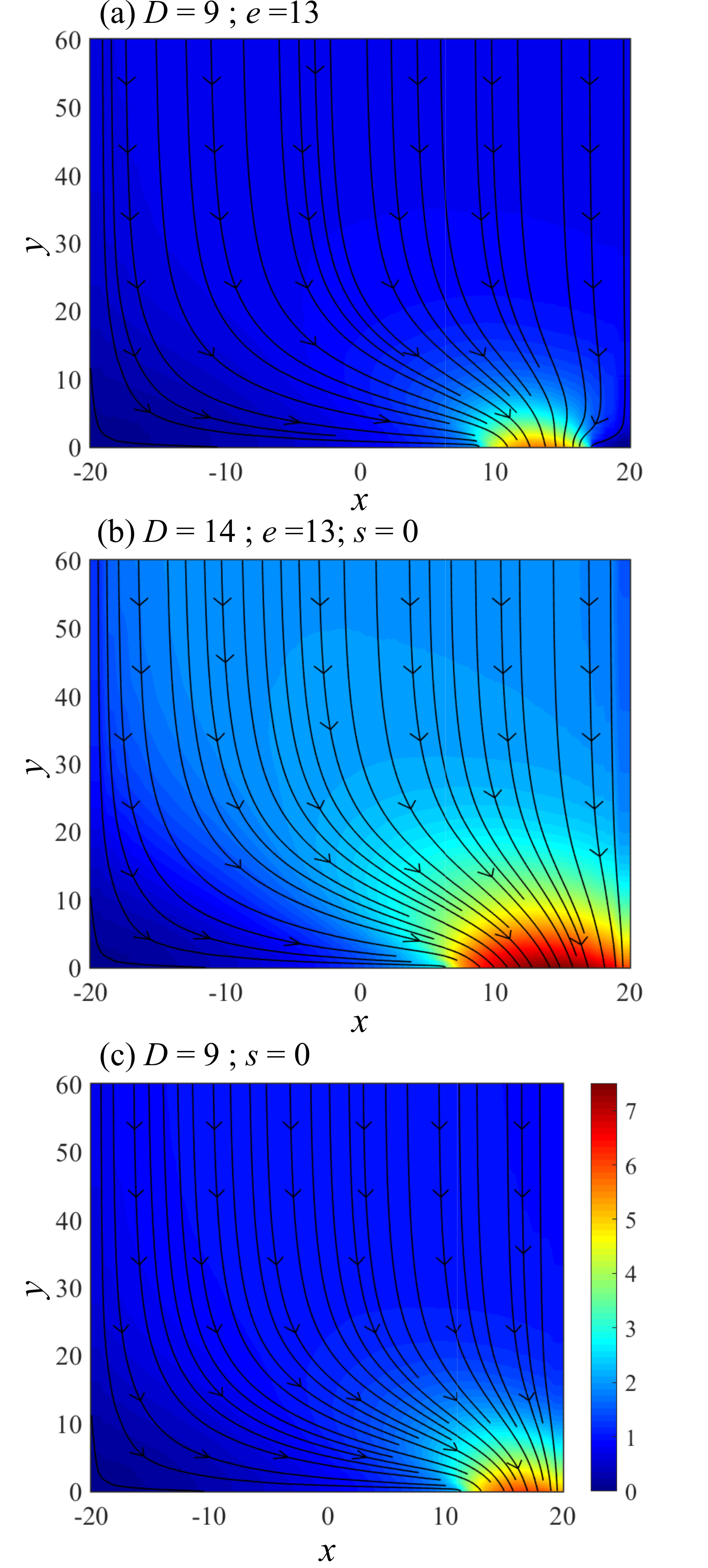}
\caption{Spatial distribution of the magnitude of mean velocity, $v = (v_x^2+v_y^2)^{1/2}$, for (a) $D=9$, $e=13$, (b) $D=14$, $e=13$ ($s=0$), and (c) $D=9$, $s=0$. Streamlines are also displayed for each case. The data are shown up to $y=60$ for clarity. Color scale for all plots is provided in (c).}
\label{fig:vspatial}
\end{figure}

Let us now proceed to the case of eccentrically-placed outlets. The variation of $v_y$ and $\overline{v}_y$ with $\hat{x}$ for $e=13$ is shown in Figs.~\ref{fig:ve13}(a) and \ref{fig:ve13}(b), respectively. It is worth remarking that the outlet of size $D=14$ touches the right wall for the said eccentricity. We note in Fig.~\ref{fig:ve13}(a) that the velocity profiles are asymmetric, with the maximum velocity occurring at the location right to the outlet center. The velocity profile for $D=14$ is noticeably distinct from the rest, wherein the velocity first grows as one travels towards the right wall from the left corner and reduces while approaching the right corner due to the presence of the wall. Further, as displayed in Fig.~\ref{fig:ve13}(b), the scaled velocity data corresponding to different $D$ do not lie on a single curve, demonstrating the breakdown of self-similar properties of velocity. This finding indicates that $e$ is not an appropriate measure of eccentricity. The parameter $e$, nevertheless, provides a geometrical consistency across the outlets of varying sizes, but it leads to comparatively differing physical conditions between them for high eccentricity, as exhibited by the streamlines and spatial distribution of the velocity field in Fig.~\ref{fig:vspatial} for the smallest ($D=9$) and the largest ($D=14$) outlets. For instance, the flow pattern is different in both the cases in the sense that grains move towards the outlet from both sides of it for $D=9$ and only from one side for $D=14$, as illustrated by the streamlines (cf. Figs.~\ref{fig:vspatial}(a) and \ref{fig:vspatial}(b)). This observation suggests that the gap between the right corners of all outlets and the right wall should be kept equal. By doing so, we get similar flow conditions for all the outlets as demonstrated for the largest and the smallest ones in Figs.~\ref{fig:vspatial}(b) and \ref{fig:vspatial}(c), respectively, for the case where the exits touch the side wall, i.e., $s=0$. The similarity between the flow fields is also maintained for other values of $s$ (not shown for brevity). Thus, $s$ may be considered as a suitable parameter for measuring the eccentricity of an outlet. 

% Figures for the velocity and its scaling
\begin{figure*}[ht!]
\includegraphics[scale=0.5]{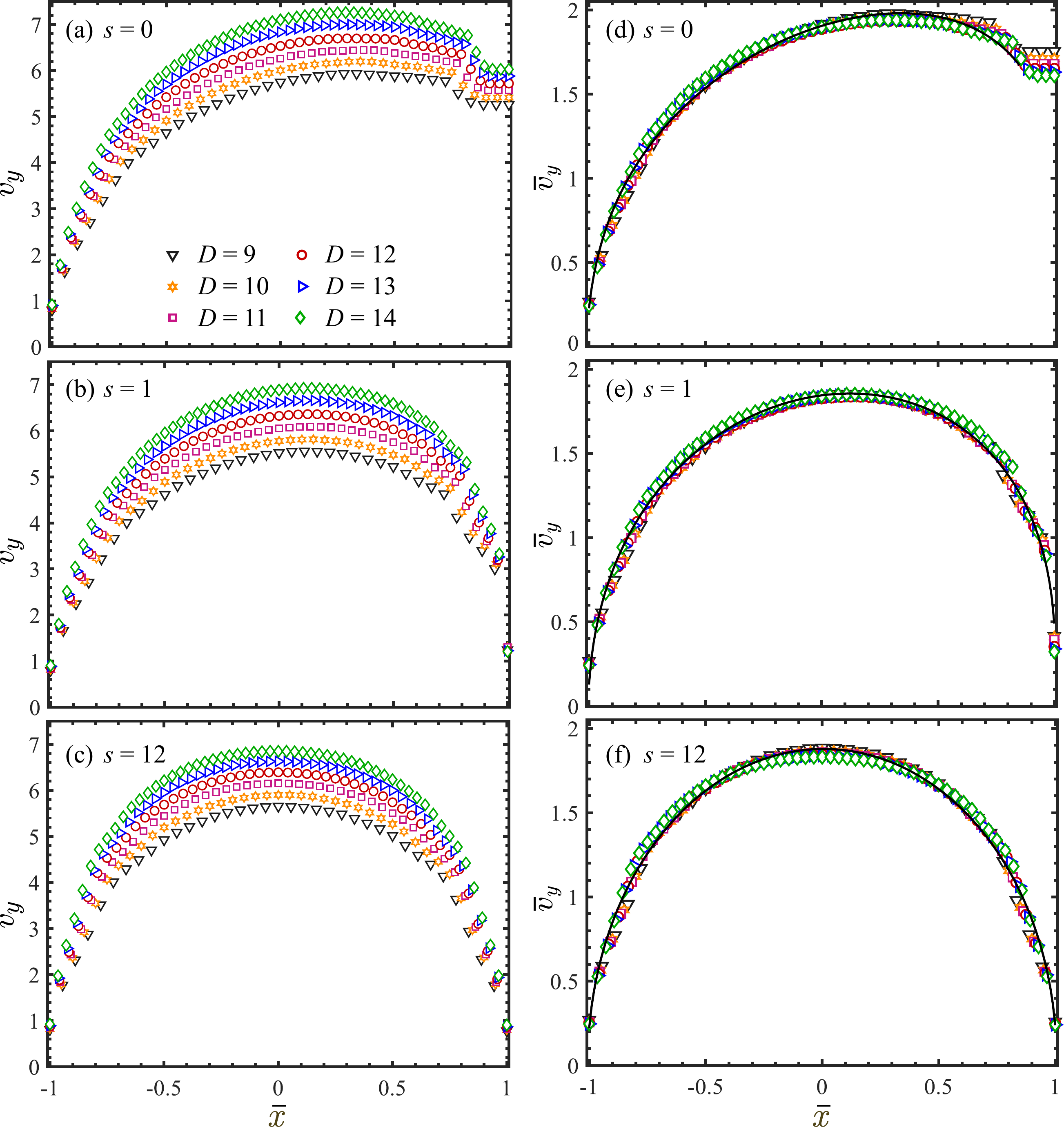}
\caption{[(a)-(c)] Horizontal profiles of the vertical component of the mean velocity $v_y$. [(d)-(f)] Variation of the scaled velocity $\overline{v}_y$ with $\overline{x}$. The data are displayed for three values of $s$, mentioned suitably on each plot. The black solid lines in (d)-(f) are fits of Eq.~(\ref{eqn:scale2}). Error bars are smaller than the markers, hence not shown. Legend for all plots is provided in (a).}
\label{fig:vedge}
\end{figure*}

Now, we consider the outlets of different sizes having equal distance between their right corners and the side wall. For brevity, we present data for three values of $s=\{0,1,12\}$, focussing on the locations close to the wall and far away from it. Fig.~\ref{fig:vedge} displays the variation of $v_y$ and $\overline{v}_y$ with $\overline{x}=2(x-e)/D$, where $e=[(W-D)/2 - s]$ is different for each outlet. Note that $\overline{x}=-1,0$ and $1$ denote the left corner, center and the right corner of an outlet, respectively. Expectedly, as Figs.~\ref{fig:vedge}(a) to \ref{fig:vedge}(c) show, the velocity rises as the aperture widens for a given $s$. The asymmetry in the velocity profiles about the outlet center ($\overline{x}=0$) is evident, which reduces as $s$ increases. The occurrence of non-zero slip velocity at the wall for $s=0$ aligns well with the results reported by Maiti \textit{et al.}\cite{maiti2016a}. Further, as displayed in Figs.~\ref{fig:vedge}(d)-\ref{fig:vedge}(f), the velocity profiles collapse onto a single curve when scaled with $D$ being the length scale. It is worth remarking that the collapse of velocity profiles is obtained for all values of $s$, unlike the previous case where the self-similarity does not hold for high values of $e$. The occurrence of such scaling for the highest eccentricity ($s=0$) is promising and extends the applicability of the self-similarity of velocity profiles in the centric silo\cite{janda2012} to the one having eccentrically-located outlets. This finding generalizes the notion of the scaling of velocity, presented in the works of Janda \textit{et al.}\cite{janda2012} and others in the centric silos, thereby providing a basis for a unified understanding of the mechanics of granular flow in proximity to an outlet.

\begin{figure}[ht!]
\includegraphics[scale=0.5]{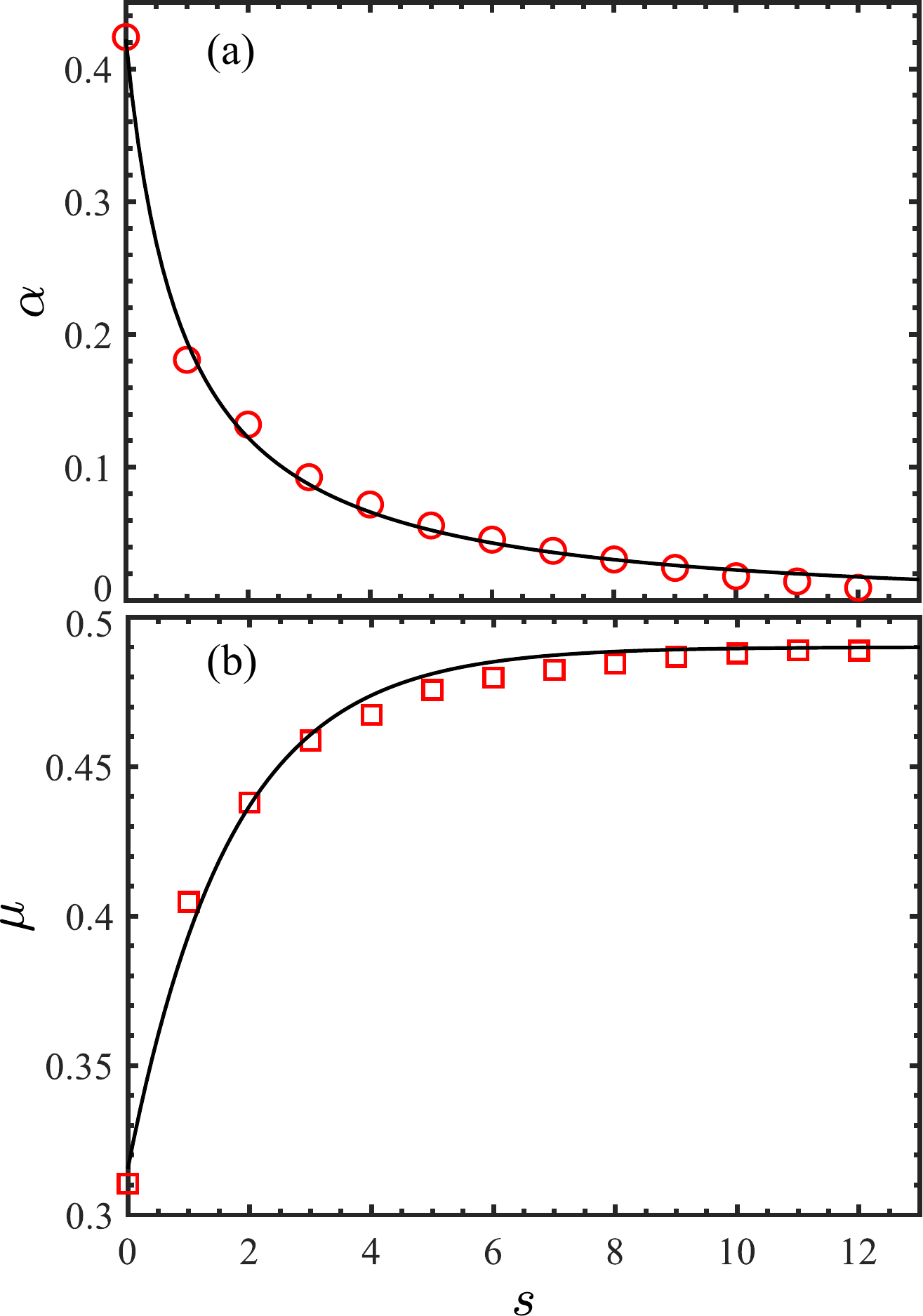}
\caption{Variation of the fitting parameters (a) $\alpha$ and (b) $\mu$ with $s$. Solid lines in (a) and (b) are fits of Eqs.~(\ref{eq:alpha}) and (\ref{eq:mu}), respectively.}
\label{fig:fit}
\end{figure}

The scaled velocity data for all $s$ are fit to the following equation
\begin{equation}
\overline{v}_y = \alpha \, \overline{x} + \beta \, (1-\gamma \, \overline{x}^2)^{\mu},
\label{eqn:scale2}
\end{equation}
which is an amended version of Eq.~(\ref{eqn:scale1}). Here, all points within the outlet are considered for fitting except for $s=0$, wherein the points lying within $1d$ distance from the wall are omitted. The fitting parameters are $\beta=1.86$ and $\gamma=0.98$, and $\alpha$ and $\mu$ are functions of $s$. Note that $\beta \approx \alpha_1$ and $\gamma = \alpha_2$ are almost invariant with $s$. Therefore, their mean values are reported. The parameter $\alpha$ in Eq.~(\ref{eqn:scale2}) models asymmetry of the velocity profile and the symmetry is recovered for $\alpha=0$. Fig.~\ref{fig:fit}(a) displays the variation of $\alpha$ with $s$. We note that $\alpha$ decreases in an exponential manner as $s$ increases and tends to vanish for higher $s$. This observation suggests the following functional form of $\alpha$, ensuring that the profile becomes more symmetric as $s$ increases,
\begin{equation}
\alpha = \frac {a}{(1+bs)} \,\, \text{exp}(-cs),
\label{eq:alpha}
\end{equation}
with $a=0.42, b=1.07$ and $c=0.046$ being the fitting parameters. The exponent $\mu$ dictates the profile shape and its variation with $s$ is shown in Fig.~\ref{fig:fit}(b). Here, $\mu$ increases exponentially and becomes nearly constant for large $s$. This is expected as velocity profiles tend to become symmetric as the outlets move away from the walls towards the center of the silo base. Thus, $\mu=\nu$, when the outlet center coincides with that of the silo base. These arguments lead to the following expression of $\mu$,
\begin{equation}
\mu = \nu - A\,\, \text{exp}(-Bs),
\label{eq:mu}
\end{equation}
where $A=0.17$ and $B=0.59$. Note that $\nu=0.49$ for the centric case, as computed earlier by fitting Eq.~(\ref{eqn:scale1}). In passing, it is worth remarking that having the exponent $\mu$ being a function of $s$ is not unexpected, given the geometry under consideration. Similarly to this finding, a recent study by Darias \textit{et al.}\cite{darias2020} reports the dependence of the exponent, controlling the shape of velocity profile, on the hopper angle.

% Figure for the variation of fitting parameters
%

%---------------------------------
% Solid fraction
%---------------------------------
\subsection{Solid fraction}
\begin{figure*}[ht!]
\includegraphics[scale=0.5]{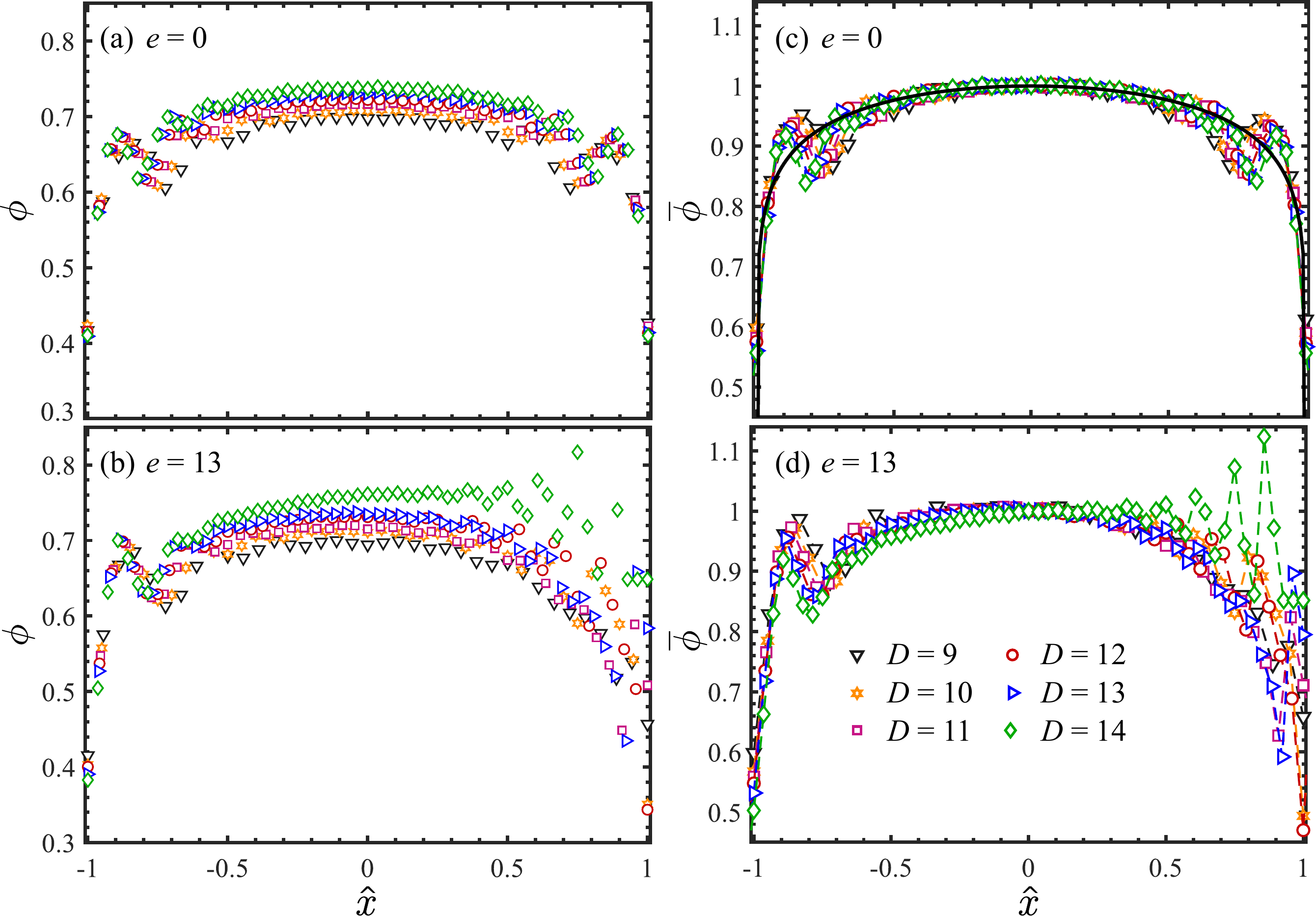}
\caption{[(a)-(b)] Variation of the solid fraction $\phi$ with $\hat{x}$ for $e=0$ and $e=13$. Error bars are smaller than markers, therefore, not plotted. [(c)-(d)] The scaled solid fraction $\overline{\phi}$ versus $\hat{x}$ for $e=0$ and $e=13$. In (c), the black solid line represents Eq.~(\ref{eq:pecc}). Legend for all plots is given in (d).}
\label{fig:pecc}
\end{figure*}

After velocity, it is natural to explore the behaviour of solid fraction $\phi$ at the exit. Considering the procedure followed for velocity, we first vary $e$ for all outlet sizes. The variation of $\phi$ for $e=0$ and $13$ with $\hat{x}=2(x-e)/D$ is shown in Figs.~\ref{fig:pecc}(a) and \ref{fig:pecc}(b), respectively. At the outset, for $e=0$, we observe that the solid fraction profiles are symmetric and the packing becomes denser near the outlet center. Further, $\phi$ rises as $D$ increases for a given horizontal location. These observations are consistent with previously reported studies\cite{janda2012,zhou2015,gella2017,mendez2021}. Asymmetry in the profiles of solid fraction for $e=13$ is clearly evident from Fig.~\ref{fig:pecc}(b) and fluctuations in $\phi$ near the right corner are due to its proximity to the wall. The horizontal profiles of scaled solid fraction $\overline{\phi} = \phi/\phi_{0}$ for $e=0$ and $13$ are displayed in Figs.~\ref{fig:pecc}(c) and \ref{fig:pecc}(d), respectively, for all outlet sizes, where $\phi_0$ is the value of solid fraction at the outlet center. For $e=0$, the self-similarity of solid fraction profiles is evident in Fig.~\ref{fig:pecc}(c) as the data collapse onto a single curve, in line with previous investigations\cite{janda2012,zhou2015,gella2017,mendez2021}. The scaled data are fit to the following expression

% Figures for the solid fraction and its scaling
\begin{equation}
\overline{\phi} = \frac{\phi}{\phi_0} = (1-\hat{x}^2)^{\eta},
\label{eq:pecc}
\end{equation}
where the fitting parameter $\eta$ governs the profile shape. Here, we obtain $\eta=0.083$, which is close to the value reported ($\eta=0.1$) in a recent study based on CFD\cite{mendez2021}. It, however, differs appreciably from those reported in experiments ($\eta=0.22$) and computations ($\eta=0.19$) of Janda \textit{et al.}\cite{janda2012} and Zhou \textit{et al.}\cite{zhou2015}, respectively. Further, the solid fraction curves do not superimpose well for the eccentricity $e=13$ as demonstrated in Fig.~\ref{fig:pecc}(d). Specifically, the profile corresponding to the largest outlet size ($D=14$) deviates from the rest of the curves near the wall, similarly to the vertical velocity (see Fig.~\ref{fig:ve13}(b)).

% Figures for the solid fraction and its scaling
\begin{figure*}[ht!]
\includegraphics[scale=0.5]{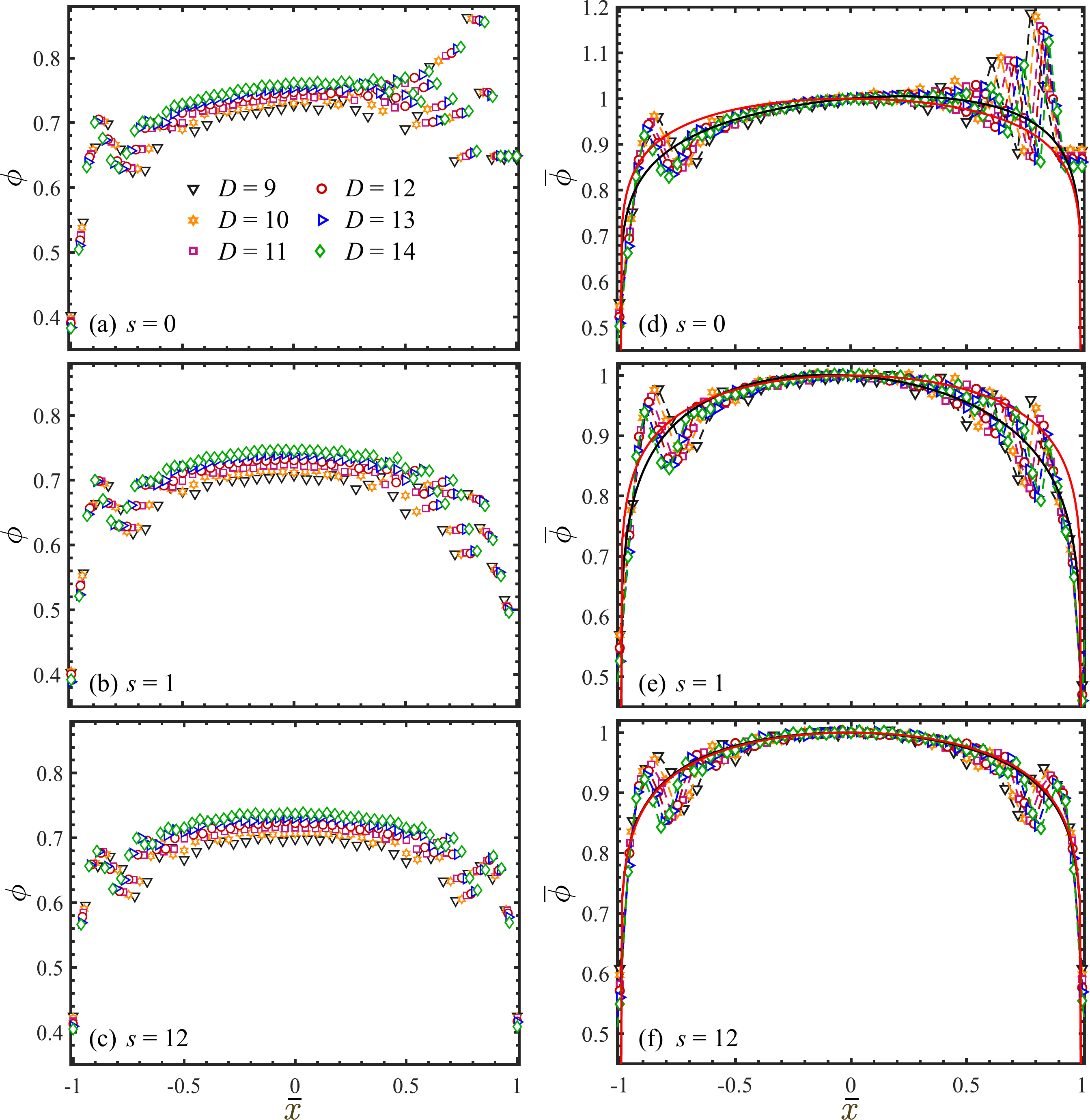}
\caption{[(a)-(c)] Variation of $\phi$ with $\overline{x}$.  Error bars are smaller than markers, hence not shown. [(d)-(f)] Horizontal profiles of the scaled solid fraction $\overline{\phi}$. The red solid lines represent Eq.~(\ref{eq:pedge}) for $C=0$ and $\xi=\eta=0.083$. The black solid lines are fits of Eq.~(\ref{eq:pedge}), which tend to overlap the red ones for high $s$, thereby demonstrating the decrease in asymmetry as $s$ increases. The data are presented for three values of $s$, mentioned appropriately on each plot.}
\label{fig:pedge}
\end{figure*}

\begin{figure}[ht!]
\includegraphics[scale=0.5]{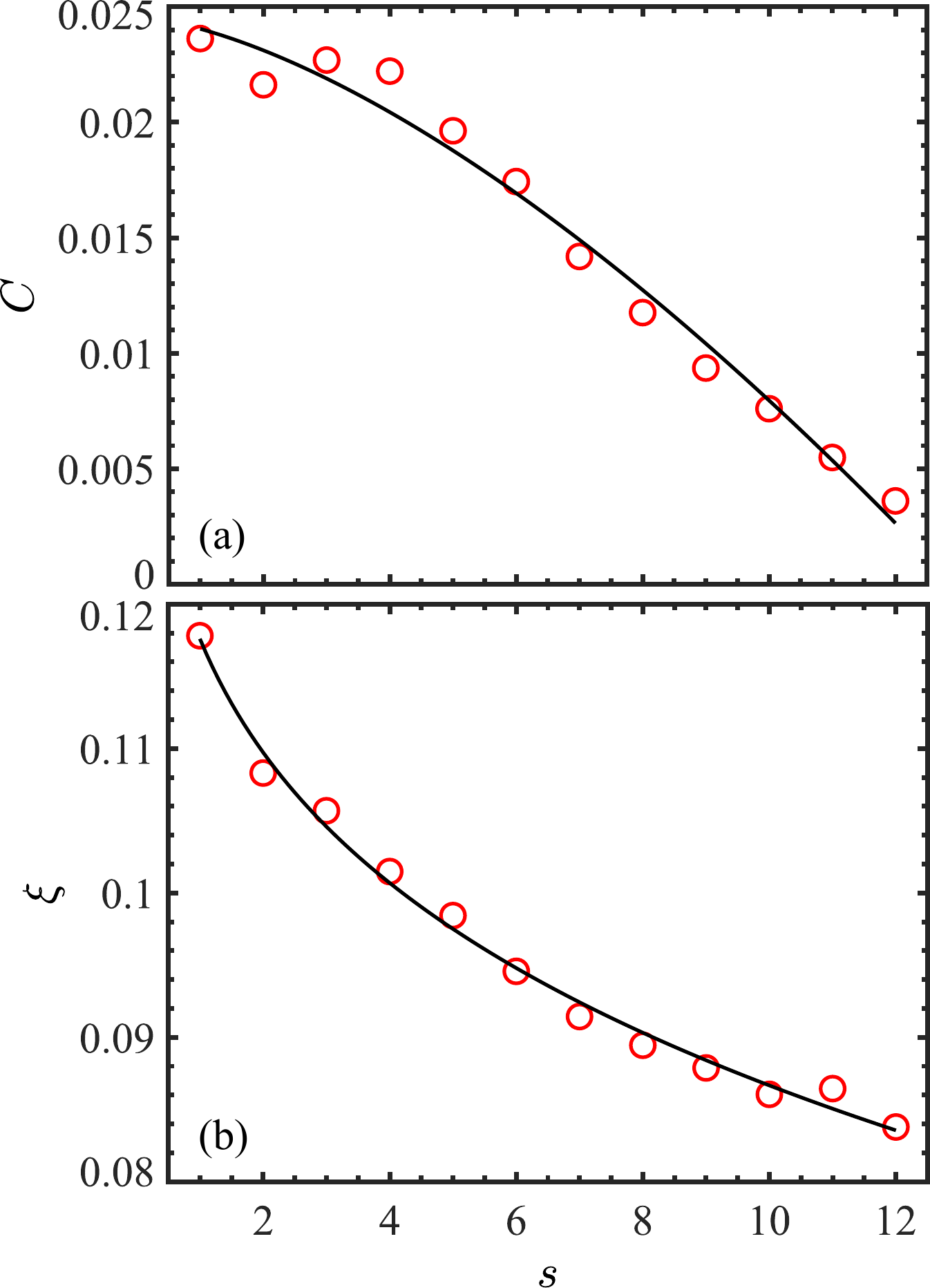}
\caption{Variation of the fitting parameters (a) $C$ and (b) $\xi$ with $s$. Solid lines in (a) and (b) are fits of Eqs.~(\ref{eq:C}) and (\ref{eq:xi}), respectively.}
\label{fig:fit-1}
\end{figure}

We now examine the behaviour of solid fraction for the scenario where the eccentricity of an outlet is given in terms of $s$. The variation of solid fraction with $\overline{x}=2(x-e)/D$ for $s=0,1$ and $12$ are presented in Figs.~\ref{fig:pedge}(a)-\ref{fig:pedge}(c), respectively, where $e=[(W-D)/2-s]$ changes with the outlet size for a given $s$. Our observations here are quite similar to what we noted earlier, including large fluctuations in $\phi$ near the wall for high eccentricities ($s=0$ and $1$) due to prominent boundary effects. Besides, we notice that the profiles become symmetric as $s$ increases. The variation of scaled solid fraction $\overline{\phi}$ with $\overline{x}$ for $s=0,1$ and $12$ is shown, respectively, in Figs.~\ref{fig:pedge}(d)-\ref{fig:pedge}(f). We note that the data collapse very well onto a single curve, except at the locations close to the wall. The scaled solid fraction data for all $s$ are fit to the following equation
\begin{equation}
\overline{\phi}=  (1-\overline{x}^2)^{\xi} \,\, \text{exp}(-C\,\overline{x}),
\label{eq:pedge}
\end{equation}
which is a modified edition of Eq.~(\ref{eq:pecc}). For fitting the data, all points within the orifice are taken into account except for $s=0$, wherein the points located upto the distance of one mean particle diameter from the side wall are excluded. The exponential factor accounts for the asymmetry in solid fraction profiles. The parameter $C$ is a function of $s$, tending to zero as $s$ increases. We obtain $C=-0.0425$ for the case when outlets touch the side wall, i.e., $s=0$, whereas $C$ is positive and varies in a non-linear fashion with $s$ ranging between 1 and 12, as shown in Fig.~\ref{fig:fit-1}. The parameter $C$ is fit to the following expression for $s=[1,12]$
\begin{equation}
C = c_1 - d_1\, s^{\lambda_1},
\label{eq:C}
\end{equation}
where $c_1=0.025, d_1 = 0.0005$ and $\lambda_1=1.5$. The exponent $\xi$ controlling the shape of solid fraction profiles also depends on $s$, and reduces as $s$ increases. The exponent $\xi = 0.089$ for $s=0$, which does not differ significantly from its centric counterpart ($\eta=0.083$). The variation of $\xi$ with $s$ is presented in Fig.~\ref{fig:fit-1}(b) for $s\geq 1$ and the data are fit to the following expression
\begin{equation}
\xi = c_2 - d_2\, s^{\lambda_2},
\label{eq:xi}
\end{equation}
with the fitting parameters being $c_2 = 0.17$, $d_2 = 0.053$ and $\lambda_2 = 0.2$. We observe in Fig.~\ref{fig:fit-1} that Eqs.~(\ref{eq:C}) and (\ref{eq:xi}) fit the respective data quite well. It is worth pointing out while closing that the dependence of $\xi$ on $s$ is expected as a result of the geometry under investigation, which aligns well with a previous work of Darias \textit{et al.}\cite{darias2020} reporting the dependence of the exponent governing the shape of solid fraction profiles on hopper angle, a pertinent geometrical variable in their case.

% Figure for the variation of fitting parameter
% Conclusions/Summary
\section{Conclusions}
\label{sec:conclude}
Utilizing discrete element simulations, we investigate the flow of granular media under gravity through eccentrically-placed exits in a two-dimensional silo. The eccentric positions of outlets are measured in terms of two parameters $s$ and $e$. The former gives the distance between the right outlet corner and the right wall, and the latter characterizes the distance between the centers of the base and exits. The profiles of vertical velocity $v_y$ and solid fraction $\phi$ are computed at the outlet, which are found to be self-similar for all values of $s$ considered. On the contrary, the self-similarity of $v_y$ and $\phi$ does not hold for all values of $e$. In view of the self-similarity of velocity and density profiles, these findings suggest that the latter measure of eccentricity, $e$, is not appropriate, which is normally used for characterizing the eccentric position of outlets in a silo. The suitability of the parameter $s$ is, nonetheless, evident, thereby providing a generalization to the findings of Janda \textit{et al.}\cite{janda2012} for centric outlets and, consequently, a basis for a unified description of the motion of discrete media near exits in other similar scenarios such as egress of pedestrians and traffic flow. Finally, the expressions for the scaled velocity and solid fraction are proposed considering the eccentricity ($s$) of outlets. The fitting parameters, including the exponent dictating the profile shape, are found to be a function of $s$, as expected. 

\section*{Acknowledgements}
A.B. gratefully acknowledges the financial support from the Indian Institute of Technology Goa through Start-up Grant (2019/SG/AB/025). A.B. thanks Dr.~Sandip Mandal, IIT (ISM) Dhanbad, for providing useful suggestions.
%
%\section*{Conflict of Interest}
%The authors have no conflicts to disclose.
%%
%\section*{Data Availability}
%The data that support the findings of this study are available from the corresponding author upon reasonable request.
%%
\bibliography{mybib.bib}	% Produces the bibliography via BibTeX.

\providecommand{\noopsort}[1]{}\providecommand{\singleletter}[1]{#1}%
\begin{thebibliography}{30}%
\makeatletter
\providecommand \@ifxundefined [1]{%
 \@ifx{#1\undefined}
}%
\providecommand \@ifnum [1]{%
 \ifnum #1\expandafter \@firstoftwo
 \else \expandafter \@secondoftwo
 \fi
}%
\providecommand \@ifx [1]{%
 \ifx #1\expandafter \@firstoftwo
 \else \expandafter \@secondoftwo
 \fi
}%
\providecommand \natexlab [1]{#1}%
\providecommand \enquote  [1]{``#1''}%
\providecommand \bibnamefont  [1]{#1}%
\providecommand \bibfnamefont [1]{#1}%
\providecommand \citenamefont [1]{#1}%
\providecommand \href@noop [0]{\@secondoftwo}%
\providecommand \href [0]{\begingroup \@sanitize@url \@href}%
\providecommand \@href[1]{\@@startlink{#1}\@@href}%
\providecommand \@@href[1]{\endgroup#1\@@endlink}%
\providecommand \@sanitize@url [0]{\catcode `\\12\catcode `\$12\catcode
  `\&12\catcode `\#12\catcode `\^12\catcode `\_12\catcode `\%12\relax}%
\providecommand \@@startlink[1]{}%
\providecommand \@@endlink[0]{}%
\providecommand \url  [0]{\begingroup\@sanitize@url \@url }%
\providecommand \@url [1]{\endgroup\@href {#1}{\urlprefix }}%
\providecommand \urlprefix  [0]{URL }%
\providecommand \Eprint [0]{\href }%
\providecommand \doibase [0]{http://dx.doi.org/}%
\providecommand \selectlanguage [0]{\@gobble}%
\providecommand \bibinfo  [0]{\@secondoftwo}%
\providecommand \bibfield  [0]{\@secondoftwo}%
\providecommand \translation [1]{[#1]}%
\providecommand \BibitemOpen [0]{}%
\providecommand \bibitemStop [0]{}%
\providecommand \bibitemNoStop [0]{.\EOS\space}%
\providecommand \EOS [0]{\spacefactor3000\relax}%
\providecommand \BibitemShut  [1]{\csname bibitem#1\endcsname}%
\let\auto@bib@innerbib\@empty
%</preamble>
\bibitem [{\citenamefont {Beverloo}, \citenamefont {Leniger},\ and\
  \citenamefont {de~Velde}(1961)}]{beverloo1961}%
  \BibitemOpen
  \bibfield  {author} {\bibinfo {author} {\bibfnamefont {W.~A.}\ \bibnamefont
  {Beverloo}}, \bibinfo {author} {\bibfnamefont {H.}~\bibnamefont {Leniger}}, \
  and\ \bibinfo {author} {\bibfnamefont {J.~V.}\ \bibnamefont {de~Velde}},\
  }\bibfield  {title} {\enquote {\bibinfo {title} {The flow of granular solids
  through orifices},}\ }\href@noop {} {\bibfield  {journal} {\bibinfo
  {journal} {Chem. Eng. Sci.}\ }\textbf {\bibinfo {volume} {15}},\ \bibinfo
  {pages} {260--269} (\bibinfo {year} {1961})}\BibitemShut {NoStop}%
\bibitem [{\citenamefont {Mankoc}\ \emph {et~al.}(2007)\citenamefont {Mankoc},
  \citenamefont {Janda}, \citenamefont {Arevalo}, \citenamefont {Pastor},
  \citenamefont {Zuriguel}, \citenamefont {Garcimart{\'\i}n},\ and\
  \citenamefont {Maza}}]{mankoc2007}%
  \BibitemOpen
  \bibfield  {author} {\bibinfo {author} {\bibfnamefont {C.}~\bibnamefont
  {Mankoc}}, \bibinfo {author} {\bibfnamefont {A.}~\bibnamefont {Janda}},
  \bibinfo {author} {\bibfnamefont {R.}~\bibnamefont {Arevalo}}, \bibinfo
  {author} {\bibfnamefont {J.~M.}\ \bibnamefont {Pastor}}, \bibinfo {author}
  {\bibfnamefont {I.}~\bibnamefont {Zuriguel}}, \bibinfo {author}
  {\bibfnamefont {A.}~\bibnamefont {Garcimart{\'\i}n}}, \ and\ \bibinfo
  {author} {\bibfnamefont {D.}~\bibnamefont {Maza}},\ }\bibfield  {title}
  {\enquote {\bibinfo {title} {The flow rate of granular materials through an
  orifice},}\ }\href@noop {} {\bibfield  {journal} {\bibinfo  {journal}
  {Granular Matter}\ }\textbf {\bibinfo {volume} {9}},\ \bibinfo {pages}
  {407--414} (\bibinfo {year} {2007})}\BibitemShut {NoStop}%
\bibitem [{\citenamefont {Janda}, \citenamefont {Zuriguel},\ and\ \citenamefont
  {Maza}(2012)}]{janda2012}%
  \BibitemOpen
  \bibfield  {author} {\bibinfo {author} {\bibfnamefont {A.}~\bibnamefont
  {Janda}}, \bibinfo {author} {\bibfnamefont {I.}~\bibnamefont {Zuriguel}}, \
  and\ \bibinfo {author} {\bibfnamefont {D.}~\bibnamefont {Maza}},\ }\bibfield
  {title} {\enquote {\bibinfo {title} {Flow rate of particles through apertures
  obtained from self-similar density and velocity profiles},}\ }\href@noop {}
  {\bibfield  {journal} {\bibinfo  {journal} {Phys. Rev. Lett.}\ }\textbf
  {\bibinfo {volume} {108}},\ \bibinfo {pages} {248001} (\bibinfo {year}
  {2012})}\BibitemShut {NoStop}%
\bibitem [{\citenamefont {Rubio-Largo}\ \emph {et~al.}(2015)\citenamefont
  {Rubio-Largo}, \citenamefont {Janda}, \citenamefont {Maza}, \citenamefont
  {Zuriguel},\ and\ \citenamefont {Hidalgo}}]{rubio2015}%
  \BibitemOpen
  \bibfield  {author} {\bibinfo {author} {\bibfnamefont {S.~M.}\ \bibnamefont
  {Rubio-Largo}}, \bibinfo {author} {\bibfnamefont {A.}~\bibnamefont {Janda}},
  \bibinfo {author} {\bibfnamefont {D.}~\bibnamefont {Maza}}, \bibinfo {author}
  {\bibfnamefont {I.}~\bibnamefont {Zuriguel}}, \ and\ \bibinfo {author}
  {\bibfnamefont {R.~C.}\ \bibnamefont {Hidalgo}},\ }\bibfield  {title}
  {\enquote {\bibinfo {title} {Disentangling the free-fall arch paradox in silo
  discharge},}\ }\href@noop {} {\bibfield  {journal} {\bibinfo  {journal}
  {Phys. Rev. Lett.}\ }\textbf {\bibinfo {volume} {114}},\ \bibinfo {pages}
  {238002} (\bibinfo {year} {2015})}\BibitemShut {NoStop}%
\bibitem [{\citenamefont {Helbing}, \citenamefont {Farkas},\ and\ \citenamefont
  {Vicsek}(2000)}]{helbing2000}%
  \BibitemOpen
  \bibfield  {author} {\bibinfo {author} {\bibfnamefont {D.}~\bibnamefont
  {Helbing}}, \bibinfo {author} {\bibfnamefont {I.}~\bibnamefont {Farkas}}, \
  and\ \bibinfo {author} {\bibfnamefont {T.}~\bibnamefont {Vicsek}},\
  }\bibfield  {title} {\enquote {\bibinfo {title} {Simulating dynamical
  features of escape panic},}\ }\href@noop {} {\bibfield  {journal} {\bibinfo
  {journal} {Nature}\ }\textbf {\bibinfo {volume} {407}},\ \bibinfo {pages}
  {487--490} (\bibinfo {year} {2000})}\BibitemShut {NoStop}%
\bibitem [{\citenamefont {Pastor}\ \emph {et~al.}(2015)\citenamefont {Pastor},
  \citenamefont {Garcimart{\'\i}n}, \citenamefont {Gago}, \citenamefont
  {Peralta}, \citenamefont {Mart{\'\i}n-G{\'o}mez}, \citenamefont {Ferrer},
  \citenamefont {Maza}, \citenamefont {Parisi}, \citenamefont {Pugnaloni},\
  and\ \citenamefont {Zuriguel}}]{pastor2015}%
  \BibitemOpen
  \bibfield  {author} {\bibinfo {author} {\bibfnamefont {J.~M.}\ \bibnamefont
  {Pastor}}, \bibinfo {author} {\bibfnamefont {A.}~\bibnamefont
  {Garcimart{\'\i}n}}, \bibinfo {author} {\bibfnamefont {P.~A.}\ \bibnamefont
  {Gago}}, \bibinfo {author} {\bibfnamefont {J.~P.}\ \bibnamefont {Peralta}},
  \bibinfo {author} {\bibfnamefont {C.}~\bibnamefont {Mart{\'\i}n-G{\'o}mez}},
  \bibinfo {author} {\bibfnamefont {L.~M.}\ \bibnamefont {Ferrer}}, \bibinfo
  {author} {\bibfnamefont {D.}~\bibnamefont {Maza}}, \bibinfo {author}
  {\bibfnamefont {D.~R.}\ \bibnamefont {Parisi}}, \bibinfo {author}
  {\bibfnamefont {L.~A.}\ \bibnamefont {Pugnaloni}}, \ and\ \bibinfo {author}
  {\bibfnamefont {I.}~\bibnamefont {Zuriguel}},\ }\bibfield  {title} {\enquote
  {\bibinfo {title} {Experimental proof of faster-is-slower in systems of
  frictional particles flowing through constrictions},}\ }\href@noop {}
  {\bibfield  {journal} {\bibinfo  {journal} {Phys. Rev. E}\ }\textbf {\bibinfo
  {volume} {92}},\ \bibinfo {pages} {062817} (\bibinfo {year}
  {2015})}\BibitemShut {NoStop}%
\bibitem [{\citenamefont {Nedderman}\ \emph {et~al.}(1982)\citenamefont
  {Nedderman}, \citenamefont {T{\"u}z{\"u}n}, \citenamefont {Savage},\ and\
  \citenamefont {Houlsby}}]{nedderman1982}%
  \BibitemOpen
  \bibfield  {author} {\bibinfo {author} {\bibfnamefont {R.~M.}\ \bibnamefont
  {Nedderman}}, \bibinfo {author} {\bibfnamefont {U.}~\bibnamefont
  {T{\"u}z{\"u}n}}, \bibinfo {author} {\bibfnamefont {S.~B.}\ \bibnamefont
  {Savage}}, \ and\ \bibinfo {author} {\bibfnamefont {G.~T.}\ \bibnamefont
  {Houlsby}},\ }\bibfield  {title} {\enquote {\bibinfo {title} {The flow of
  granular materials—i: Discharge rates from hoppers},}\ }\href@noop {}
  {\bibfield  {journal} {\bibinfo  {journal} {Chem. Eng. Sci.}\ }\textbf
  {\bibinfo {volume} {37}},\ \bibinfo {pages} {1597--1609} (\bibinfo {year}
  {1982})}\BibitemShut {NoStop}%
\bibitem [{\citenamefont {Saleh}, \citenamefont {Golshan},\ and\ \citenamefont
  {Zarghami}(2018)}]{saleh2018}%
  \BibitemOpen
  \bibfield  {author} {\bibinfo {author} {\bibfnamefont {K.}~\bibnamefont
  {Saleh}}, \bibinfo {author} {\bibfnamefont {S.}~\bibnamefont {Golshan}}, \
  and\ \bibinfo {author} {\bibfnamefont {R.}~\bibnamefont {Zarghami}},\
  }\bibfield  {title} {\enquote {\bibinfo {title} {A review on gravity flow of
  free-flowing granular solids in silos--basics and practical aspects},}\
  }\href@noop {} {\bibfield  {journal} {\bibinfo  {journal} {Chem. Eng. Sci.}\
  }\textbf {\bibinfo {volume} {192}},\ \bibinfo {pages} {1011--1035} (\bibinfo
  {year} {2018})}\BibitemShut {NoStop}%
\bibitem [{\citenamefont {Sielamowicz}, \citenamefont {Czech},\ and\
  \citenamefont {Kowalewski}(2011)}]{sielamowicz2011}%
  \BibitemOpen
  \bibfield  {author} {\bibinfo {author} {\bibfnamefont {I.}~\bibnamefont
  {Sielamowicz}}, \bibinfo {author} {\bibfnamefont {M.}~\bibnamefont {Czech}},
  \ and\ \bibinfo {author} {\bibfnamefont {T.~A.}\ \bibnamefont {Kowalewski}},\
  }\bibfield  {title} {\enquote {\bibinfo {title} {Empirical analysis of
  eccentric flow registered by the dpiv technique inside a silo model},}\
  }\href@noop {} {\bibfield  {journal} {\bibinfo  {journal} {Powder Tech.}\
  }\textbf {\bibinfo {volume} {212}},\ \bibinfo {pages} {38--56} (\bibinfo
  {year} {2011})}\BibitemShut {NoStop}%
\bibitem [{\citenamefont {Sielamowicz}, \citenamefont {Czech},\ and\
  \citenamefont {Kowalewski}(2010)}]{sielamowicz2010}%
  \BibitemOpen
  \bibfield  {author} {\bibinfo {author} {\bibfnamefont {I.}~\bibnamefont
  {Sielamowicz}}, \bibinfo {author} {\bibfnamefont {M.}~\bibnamefont {Czech}},
  \ and\ \bibinfo {author} {\bibfnamefont {T.~A.}\ \bibnamefont {Kowalewski}},\
  }\bibfield  {title} {\enquote {\bibinfo {title} {Empirical description of
  flow parameters in eccentric flow inside a silo model},}\ }\href@noop {}
  {\bibfield  {journal} {\bibinfo  {journal} {Powder Tech.}\ }\textbf {\bibinfo
  {volume} {198}},\ \bibinfo {pages} {381--394} (\bibinfo {year}
  {2010})}\BibitemShut {NoStop}%
\bibitem [{\citenamefont {Maiti}\ \emph {et~al.}(2016)\citenamefont {Maiti},
  \citenamefont {Meena}, \citenamefont {Das},\ and\ \citenamefont
  {Das}}]{maiti2016a}%
  \BibitemOpen
  \bibfield  {author} {\bibinfo {author} {\bibfnamefont {R.}~\bibnamefont
  {Maiti}}, \bibinfo {author} {\bibfnamefont {S.}~\bibnamefont {Meena}},
  \bibinfo {author} {\bibfnamefont {P.~K.}\ \bibnamefont {Das}}, \ and\
  \bibinfo {author} {\bibfnamefont {G.}~\bibnamefont {Das}},\ }\bibfield
  {title} {\enquote {\bibinfo {title} {Flow field during eccentric discharge
  from quasi-two-dimensional silos--extension of the kinematic model with
  validation},}\ }\href@noop {} {\bibfield  {journal} {\bibinfo  {journal}
  {AIChE J.}\ }\textbf {\bibinfo {volume} {62}},\ \bibinfo {pages} {1439--1453}
  (\bibinfo {year} {2016})}\BibitemShut {NoStop}%
\bibitem [{\citenamefont {Nedderman}\ and\ \citenamefont
  {T{\"u}z{\"u}n}(1979)}]{nedderman1979}%
  \BibitemOpen
  \bibfield  {author} {\bibinfo {author} {\bibfnamefont {R.~M.}\ \bibnamefont
  {Nedderman}}\ and\ \bibinfo {author} {\bibfnamefont {U.}~\bibnamefont
  {T{\"u}z{\"u}n}},\ }\bibfield  {title} {\enquote {\bibinfo {title} {A
  kinematic model for the flow of granular materials},}\ }\href@noop {}
  {\bibfield  {journal} {\bibinfo  {journal} {Powder Technol.}\ }\textbf
  {\bibinfo {volume} {22}},\ \bibinfo {pages} {243--253} (\bibinfo {year}
  {1979})}\BibitemShut {NoStop}%
\bibitem [{\citenamefont {Maiti}, \citenamefont {Das},\ and\ \citenamefont
  {Das}(2016)}]{maiti2016b}%
  \BibitemOpen
  \bibfield  {author} {\bibinfo {author} {\bibfnamefont {R.}~\bibnamefont
  {Maiti}}, \bibinfo {author} {\bibfnamefont {G.}~\bibnamefont {Das}}, \ and\
  \bibinfo {author} {\bibfnamefont {P.~K.}\ \bibnamefont {Das}},\ }\bibfield
  {title} {\enquote {\bibinfo {title} {Experiments on eccentric granular
  discharge from a quasi-two-dimensional silo},}\ }\href@noop {} {\bibfield
  {journal} {\bibinfo  {journal} {Powder Tech.}\ }\textbf {\bibinfo {volume}
  {301}},\ \bibinfo {pages} {1054--1066} (\bibinfo {year} {2016})}\BibitemShut
  {NoStop}%
\bibitem [{\citenamefont {Zhou}, \citenamefont {Ruyer},\ and\ \citenamefont
  {Aussillous}(2015)}]{zhou2015}%
  \BibitemOpen
  \bibfield  {author} {\bibinfo {author} {\bibfnamefont {Y.}~\bibnamefont
  {Zhou}}, \bibinfo {author} {\bibfnamefont {P.}~\bibnamefont {Ruyer}}, \ and\
  \bibinfo {author} {\bibfnamefont {P.}~\bibnamefont {Aussillous}},\ }\bibfield
   {title} {\enquote {\bibinfo {title} {Discharge flow of a bidisperse granular
  media from a silo: {D}iscrete particle simulations},}\ }\href@noop {}
  {\bibfield  {journal} {\bibinfo  {journal} {Phys. Rev. E}\ }\textbf {\bibinfo
  {volume} {92}},\ \bibinfo {pages} {062204} (\bibinfo {year}
  {2015})}\BibitemShut {NoStop}%
\bibitem [{\citenamefont {Rubio-Largo}, \citenamefont {Maza},\ and\
  \citenamefont {Hidalgo}(2017)}]{rubio2017}%
  \BibitemOpen
  \bibfield  {author} {\bibinfo {author} {\bibfnamefont {S.~M.}\ \bibnamefont
  {Rubio-Largo}}, \bibinfo {author} {\bibfnamefont {D.}~\bibnamefont {Maza}}, \
  and\ \bibinfo {author} {\bibfnamefont {R.~C.}\ \bibnamefont {Hidalgo}},\
  }\bibfield  {title} {\enquote {\bibinfo {title} {Large-scale numerical
  simulations of polydisperse particle flow in a silo},}\ }\href@noop {}
  {\bibfield  {journal} {\bibinfo  {journal} {Comp. Part. Mech.}\ }\textbf
  {\bibinfo {volume} {4}},\ \bibinfo {pages} {419--427} (\bibinfo {year}
  {2017})}\BibitemShut {NoStop}%
\bibitem [{\citenamefont {Madrid}, \citenamefont {Asencio},\ and\ \citenamefont
  {Maza}(2017)}]{madrid2017}%
  \BibitemOpen
  \bibfield  {author} {\bibinfo {author} {\bibfnamefont {M.}~\bibnamefont
  {Madrid}}, \bibinfo {author} {\bibfnamefont {K.}~\bibnamefont {Asencio}}, \
  and\ \bibinfo {author} {\bibfnamefont {D.}~\bibnamefont {Maza}},\ }\bibfield
  {title} {\enquote {\bibinfo {title} {Silo discharge of binary granular
  mixtures},}\ }\href@noop {} {\bibfield  {journal} {\bibinfo  {journal} {Phys.
  Rev. E}\ }\textbf {\bibinfo {volume} {96}},\ \bibinfo {pages} {022904}
  (\bibinfo {year} {2017})}\BibitemShut {NoStop}%
\bibitem [{\citenamefont {Gella}, \citenamefont {Maza},\ and\ \citenamefont
  {Zuriguel}(2017)}]{gella2017}%
  \BibitemOpen
  \bibfield  {author} {\bibinfo {author} {\bibfnamefont {D.}~\bibnamefont
  {Gella}}, \bibinfo {author} {\bibfnamefont {D.}~\bibnamefont {Maza}}, \ and\
  \bibinfo {author} {\bibfnamefont {I.}~\bibnamefont {Zuriguel}},\ }\bibfield
  {title} {\enquote {\bibinfo {title} {Role of particle size in the kinematic
  properties of silo flow},}\ }\href@noop {} {\bibfield  {journal} {\bibinfo
  {journal} {Phys. Rev. E}\ }\textbf {\bibinfo {volume} {95}},\ \bibinfo
  {pages} {052904} (\bibinfo {year} {2017})}\BibitemShut {NoStop}%
\bibitem [{\citenamefont {Bhateja}(2020)}]{bhateja2020}%
  \BibitemOpen
  \bibfield  {author} {\bibinfo {author} {\bibfnamefont {A.}~\bibnamefont
  {Bhateja}},\ }\bibfield  {title} {\enquote {\bibinfo {title} {Velocity
  scaling in the region of orifice influence in silo draining under gravity},}\
  }\href@noop {} {\bibfield  {journal} {\bibinfo  {journal} {Phys. Rev. E}\
  }\textbf {\bibinfo {volume} {102}},\ \bibinfo {pages} {042904} (\bibinfo
  {year} {2020})}\BibitemShut {NoStop}%
\bibitem [{\citenamefont {Radjai}\ and\ \citenamefont
  {Richefeu}(2009)}]{radjai2009}%
  \BibitemOpen
  \bibfield  {author} {\bibinfo {author} {\bibfnamefont {F.}~\bibnamefont
  {Radjai}}\ and\ \bibinfo {author} {\bibfnamefont {V.}~\bibnamefont
  {Richefeu}},\ }\bibfield  {title} {\enquote {\bibinfo {title} {Contact
  dynamics as a nonsmooth discrete element method},}\ }\href@noop {} {\bibfield
   {journal} {\bibinfo  {journal} {Mech. Mater.}\ }\textbf {\bibinfo {volume}
  {41}},\ \bibinfo {pages} {715--728} (\bibinfo {year} {2009})}\BibitemShut
  {NoStop}%
\bibitem [{\citenamefont {Darias}\ \emph {et~al.}(2020)\citenamefont {Darias},
  \citenamefont {Gella}, \citenamefont {Fern{\'a}ndez}, \citenamefont
  {Zuriguel},\ and\ \citenamefont {Maza}}]{darias2020}%
  \BibitemOpen
  \bibfield  {author} {\bibinfo {author} {\bibfnamefont {J.~R.}\ \bibnamefont
  {Darias}}, \bibinfo {author} {\bibfnamefont {D.}~\bibnamefont {Gella}},
  \bibinfo {author} {\bibfnamefont {M.~E.}\ \bibnamefont {Fern{\'a}ndez}},
  \bibinfo {author} {\bibfnamefont {I.}~\bibnamefont {Zuriguel}}, \ and\
  \bibinfo {author} {\bibfnamefont {D.}~\bibnamefont {Maza}},\ }\bibfield
  {title} {\enquote {\bibinfo {title} {The hopper angle role on the velocity
  and solid-fraction profiles at the outlet of silos},}\ }\href@noop {}
  {\bibfield  {journal} {\bibinfo  {journal} {Powder Technol.}\ }\textbf
  {\bibinfo {volume} {366}},\ \bibinfo {pages} {488--496} (\bibinfo {year}
  {2020})}\BibitemShut {NoStop}%
\bibitem [{\citenamefont {M{\'e}ndez}, \citenamefont {Hidalgo},\ and\
  \citenamefont {Maza}(2021)}]{mendez2021}%
  \BibitemOpen
  \bibfield  {author} {\bibinfo {author} {\bibfnamefont {D.}~\bibnamefont
  {M{\'e}ndez}}, \bibinfo {author} {\bibfnamefont {R.~C.}\ \bibnamefont
  {Hidalgo}}, \ and\ \bibinfo {author} {\bibfnamefont {D.}~\bibnamefont
  {Maza}},\ }\bibfield  {title} {\enquote {\bibinfo {title} {The role of the
  hopper angle in silos: experimental and cfd analysis},}\ }\href@noop {}
  {\bibfield  {journal} {\bibinfo  {journal} {Granul. Matter}\ }\textbf
  {\bibinfo {volume} {23}},\ \bibinfo {pages} {1--13} (\bibinfo {year}
  {2021})}\BibitemShut {NoStop}%
\bibitem [{\citenamefont {Cundall}\ and\ \citenamefont
  {Strack}(1979)}]{cundall1979}%
  \BibitemOpen
  \bibfield  {author} {\bibinfo {author} {\bibfnamefont {P.~A.}\ \bibnamefont
  {Cundall}}\ and\ \bibinfo {author} {\bibfnamefont {O.~D.~L.}\ \bibnamefont
  {Strack}},\ }\bibfield  {title} {\enquote {\bibinfo {title} {A discrete
  numerical model for granular assemblies},}\ }\href@noop {} {\bibfield
  {journal} {\bibinfo  {journal} {Geotechnique}\ }\textbf {\bibinfo {volume}
  {29(1)}},\ \bibinfo {pages} {47--65} (\bibinfo {year} {1979})}\BibitemShut
  {NoStop}%
\bibitem [{\citenamefont {Sh{\"a}fer}, \citenamefont {Dippel},\ and\
  \citenamefont {Wolf}(1996)}]{shafer1996}%
  \BibitemOpen
  \bibfield  {author} {\bibinfo {author} {\bibfnamefont {J.}~\bibnamefont
  {Sh{\"a}fer}}, \bibinfo {author} {\bibfnamefont {S.}~\bibnamefont {Dippel}},
  \ and\ \bibinfo {author} {\bibfnamefont {D.~E.}\ \bibnamefont {Wolf}},\
  }\bibfield  {title} {\enquote {\bibinfo {title} {Force schemes in simulations
  of granular materials},}\ }\href@noop {} {\bibfield  {journal} {\bibinfo
  {journal} {Journal de Physique I}\ }\textbf {\bibinfo {volume} {6}},\
  \bibinfo {pages} {5--20} (\bibinfo {year} {1996})}\BibitemShut {NoStop}%
\bibitem [{\citenamefont {Mishra}(2003)}]{bkm2003}%
  \BibitemOpen
  \bibfield  {author} {\bibinfo {author} {\bibfnamefont {B.~K.}\ \bibnamefont
  {Mishra}},\ }\bibfield  {title} {\enquote {\bibinfo {title} {A review of
  computer simulation of tumbling mills by the discrete element method: Part
  {I}-contact mechanics},}\ }\href@noop {} {\bibfield  {journal} {\bibinfo
  {journal} {Int. J. Miner. Process.}\ }\textbf {\bibinfo {volume} {71}},\
  \bibinfo {pages} {73--93} (\bibinfo {year} {2003})}\BibitemShut {NoStop}%
\bibitem [{\citenamefont {Zhang}\ and\ \citenamefont
  {Whiten}(1996)}]{zhang1996}%
  \BibitemOpen
  \bibfield  {author} {\bibinfo {author} {\bibfnamefont {D.}~\bibnamefont
  {Zhang}}\ and\ \bibinfo {author} {\bibfnamefont {W.~J.}\ \bibnamefont
  {Whiten}},\ }\bibfield  {title} {\enquote {\bibinfo {title} {The calculation
  of contact forces between particles using spring and damping models},}\
  }\href@noop {} {\bibfield  {journal} {\bibinfo  {journal} {Powder
  Technology}\ }\textbf {\bibinfo {volume} {88}},\ \bibinfo {pages} {59--64}
  (\bibinfo {year} {1996})}\BibitemShut {NoStop}%
\bibitem [{\citenamefont {Kruggel-Emden}\ \emph {et~al.}(2008)\citenamefont
  {Kruggel-Emden}, \citenamefont {Sturm}, \citenamefont {Wirtz},\ and\
  \citenamefont {Scherer}}]{kruggel2008b}%
  \BibitemOpen
  \bibfield  {author} {\bibinfo {author} {\bibfnamefont {H.}~\bibnamefont
  {Kruggel-Emden}}, \bibinfo {author} {\bibfnamefont {M.}~\bibnamefont
  {Sturm}}, \bibinfo {author} {\bibfnamefont {S.}~\bibnamefont {Wirtz}}, \ and\
  \bibinfo {author} {\bibfnamefont {V.}~\bibnamefont {Scherer}},\ }\bibfield
  {title} {\enquote {\bibinfo {title} {Selection of an appropriate time
  integration scheme for the discrete element method ({DEM})},}\ }\href@noop {}
  {\bibfield  {journal} {\bibinfo  {journal} {Comput. Chem. Eng.}\ }\textbf
  {\bibinfo {volume} {32}},\ \bibinfo {pages} {2263--2279} (\bibinfo {year}
  {2008})}\BibitemShut {NoStop}%
\bibitem [{\citenamefont {Janda}\ \emph {et~al.}(2008)\citenamefont {Janda},
  \citenamefont {Zuriguel}, \citenamefont {Garcimart{\'\i}n}, \citenamefont
  {Pugnaloni},\ and\ \citenamefont {Maza}}]{janda2008}%
  \BibitemOpen
  \bibfield  {author} {\bibinfo {author} {\bibfnamefont {A.}~\bibnamefont
  {Janda}}, \bibinfo {author} {\bibfnamefont {I.}~\bibnamefont {Zuriguel}},
  \bibinfo {author} {\bibfnamefont {A.}~\bibnamefont {Garcimart{\'\i}n}},
  \bibinfo {author} {\bibfnamefont {L.~A.}\ \bibnamefont {Pugnaloni}}, \ and\
  \bibinfo {author} {\bibfnamefont {D.}~\bibnamefont {Maza}},\ }\bibfield
  {title} {\enquote {\bibinfo {title} {Jamming and critical outlet size in the
  discharge of a two-dimensional silo},}\ }\href@noop {} {\bibfield  {journal}
  {\bibinfo  {journal} {EPL}\ }\textbf {\bibinfo {volume} {84}},\ \bibinfo
  {pages} {44002} (\bibinfo {year} {2008})}\BibitemShut {NoStop}%
\bibitem [{\citenamefont {Kondic}(2014)}]{kondic2014}%
  \BibitemOpen
  \bibfield  {author} {\bibinfo {author} {\bibfnamefont {L.}~\bibnamefont
  {Kondic}},\ }\bibfield  {title} {\enquote {\bibinfo {title} {Simulations of
  two dimensional hopper flow},}\ }\href@noop {} {\bibfield  {journal}
  {\bibinfo  {journal} {Granular Matter}\ }\textbf {\bibinfo {volume} {16}},\
  \bibinfo {pages} {235--242} (\bibinfo {year} {2014})}\BibitemShut {NoStop}%
\bibitem [{\citenamefont {Goldhirsch}(2010)}]{goldhirsch2010}%
  \BibitemOpen
  \bibfield  {author} {\bibinfo {author} {\bibfnamefont {I.}~\bibnamefont
  {Goldhirsch}},\ }\bibfield  {title} {\enquote {\bibinfo {title} {Stress,
  stress asymmetry and couple stress: from discrete particles to continuous
  fields},}\ }\href@noop {} {\bibfield  {journal} {\bibinfo  {journal}
  {Granular Matter}\ }\textbf {\bibinfo {volume} {12}},\ \bibinfo {pages}
  {239--252} (\bibinfo {year} {2010})}\BibitemShut {NoStop}%
\bibitem [{\citenamefont {Altman}\ and\ \citenamefont
  {Bland}(2005)}]{altman2005}%
  \BibitemOpen
  \bibfield  {author} {\bibinfo {author} {\bibfnamefont {D.~G.}\ \bibnamefont
  {Altman}}\ and\ \bibinfo {author} {\bibfnamefont {J.~M.}\ \bibnamefont
  {Bland}},\ }\bibfield  {title} {\enquote {\bibinfo {title} {Standard
  deviations and standard errors},}\ }\href@noop {} {\bibfield  {journal}
  {\bibinfo  {journal} {BMJ}\ }\textbf {\bibinfo {volume} {331}},\ \bibinfo
  {pages} {903} (\bibinfo {year} {2005})}\BibitemShut {NoStop}%
\end{thebibliography}%

% End document
\end{document}